\documentclass[twocolumn,showpacs,prb,10pt]{revtex4}
\usepackage[english]{babel}
\usepackage{amsmath,amssymb} 
\usepackage{times}
\usepackage{epsfig} 

\begin{document} 
\title{Finite-temperature dynamics with the density-matrix
renormalization group method}

\author{J. Kokalj$^1$ and P. Prelov\v sek$^{1,2}$}
\affiliation{$^1$J.\ Stefan Institute, SI-1000 Ljubljana, Slovenia}
\affiliation{$^2$ Faculty of Mathematics and Physics, University of
Ljubljana, SI-1000 Ljubljana, Slovenia}

\date{\today}

\begin{abstract}
We present a new numerical method for the evaluation of dynamical
response functions at finite temperatures in one-dimensional strongly
correlated systems.  The approach is based on the density-matrix
renormalization group method, combined with the finite-temperature
Lanczos diagonalization.  The feasibility of the method is
tested on the example of dynamical spin correlations in the
anisotropic Heisenberg chain, in particular it yields nontrivial
results for the critical behavior in the isotropic case.
\end{abstract}

\maketitle 

\section{Introduction}

Strongly correlated systems present one of the major theoretical
challenges in last decades and are stimulating the intensive search
for adequate numerical methods to evaluate their properties. Within
the low-dimensional systems, in particular one-dimensional (1D)
systems the breakthrough has been achieved with the introduction of
the density matrix renormalization group (DMRG) method \cite{white92}
allowing accurate calculation of the ground-state wavefunction and its
static properties on large systems far beyond those available with the
exact-diagonalization methods.  Among various DMRG extensions
\cite{schollwock05} we concentrate here on the goal to study the
dynamical response of such systems at finite temperatures $T>0$. It
should be observed that in spite of the satisfactory description and
an understanding of static properties of generic 1D systems at
$T>0$ the corresponding dynamics, in particular the low-frequency one
as manifested in the transport quantities, NMR relaxation, is far
less understood and approachable via numerical methods.

For dynamical response within the ground-state the targeting within
the DMRG has been extended to contain also excited
states.\cite{hallberg95,kuhner99} Transfer-matrix DMRG
\cite{nishino95,bursill96,naef99} is very efficient to evaluate
thermodynamic properties of models with short-range interactions, as
well as some dynamical correlations of very limited range. Time
dependent DMRG \cite{cazalilla02,zwolak04} developed recently enables
studies of short-time evolution of general many-body systems, hence
also of $T>0$ behavior, but is rather limited in reaching the
low-$\omega$ response. Recently, a DMRG method extended with the
polynomial expansion has been proposed to treat low-$T$
dynamics.\cite{sota08} On the other hand, methods emerging from the
exact diagonalization approach as the $T>0$ Lanczos method (FTLM)
\cite{jaklic94} and the low-$T$ version \cite{aichhorn03} have
high-$\omega$ resolution and provide the information on the nontrivial
dynamics of correlated models, but are still restricted to small
systems reachable with exact diagonalization.

The paper is organized as follows.  In the next section
(\ref{sec:method}) we first present our new method, with which we
calculate some static and dynamical properties of the model described
in section \ref{sec:model}. In section \ref{sec:model} we also show
our results, first the test of our method on the $XY$ model and then
our main results for the isotropic Heisenberg model. In the last
section \ref{sec:conclusions} we present our conclusions.

\section{Method}
\label{sec:method}
In this paper we propose a new method for the calculation of the
$T>0$ dynamics which is a combination of the FTLM and the DMRG, namely
the finite-temperature dynamical DMRG (FTD-DMRG) method.  It is
constructed to calculate dynamical response functions in 1D systems at
$T>0$, with the emphasis on the low-$\omega$ regime. As a test we
consider highly nontrivial spin correlations within the anisotropic
Heisenberg model on a chain.

In the standard $T=0$ DMRG the ground-state is used to
construct the basis. In our case we use the full $T>0$ density matrix,
which can in general be expressed with
eigenstates $|n\rangle$ and corresponding eigenvalues $E_n$,
\begin{equation}
\hat \rho =\frac{1}{Z} e^{-\beta \hat H}=
\frac{1}{Z} \sum_{n=1}^{N_{st}} |n\rangle e^{-\beta E_n} \langle n| ,
\label{rho}
\end{equation}
where $\beta=1/T$ and $Z$ is the (grand)canonical sum.
We proceed by extending the density matrix, Eq.(\ref{rho}), with the sampling over
the random vectors $|r\rangle = \sum_n \beta_{rn} |n\rangle$ where
$\beta_{rn}$ denote random amplitudes,
\begin{equation}
\hat \rho \sim \frac{N_{st}}{Z R} \sum_{r=1}^R e^{-\beta \hat H/2}|r\rangle
  \langle r|  e^{-\beta  \hat H/2}.
\label{rho2}  
\end{equation}
It is easy to show that Eq.(\ref{rho2}) reduces to Eq.(\ref{rho})
expressed in diagonal basis $|n\rangle\langle n|$ since
offdiagonal terms vanish assuming normalized and random $|r\rangle$.
\cite{jaklic94}

In Eq. (\ref{rho2}) we evaluate the operator $e^{-\beta \hat H/2}$ on
$|r\rangle$ by starting the Lanczos procedure from $|r\rangle$.  After
diagonalization of the Lanczos tridiagonal $\hat H$, we obtain the first
series of Lanczos eigenvectors $|\psi_i^r\rangle$ with corresponding
eigenenergies $\epsilon_i^r$,
\begin{eqnarray}
|\tilde \psi_r\rangle &=&\sum_{i=1}^Me^{-\beta \epsilon^r_i/2}|\psi_i^r
\rangle \langle \psi_i^r |r\rangle, \nonumber \\
\hat \rho &\sim& \frac{N_{st}}{Z R} \sum_{r=1}^R |\tilde \psi_r \rangle
\langle \tilde \psi_r |.
\label{rho3}
\end{eqnarray}
It is evident that for $M$ approaching $N_{st}$ Eq.~(\ref{rho3})
reproduces fully Eq.~(\ref{rho}), while for $M \ll N_{st}$ as used in
practice represents an efficient way of evaluation of density matrix.
The sum $Z$ may be evaluated in the same manner as within the FTLM
\cite{jaklic94}
\begin{equation}
Z \sim \frac{N_{st}}{R}\sum_{r=1}^R\sum_{i=1}^M e^{-\beta
  \epsilon_i^r}|\langle\psi_i^r|r\rangle|^2 .
\end{equation}
In the original $T=0$ DMRG procedure one targets the
ground-state. \cite{white92,schollwock05} Instead, at $T>0$ we target states
$|\tilde \psi_r\rangle$ and construct the density matrix according to
Eq.~(\ref{rho3}).

Since our aim is to calculate dynamical response
functions, expressed as autocorrelation functions, we also require a
good representation of the operator density matrix,
\begin{equation}
\hat \rho_A =\frac{1}{Z} \sum_{n=1}^{N_{st}} |\hat A n\rangle
e^{-\beta E_n}  \langle \hat A n|.
\label{rhoa}
\end{equation}
It replaces the operator on the ground-state in original $T=0$ DMRG \cite{schollwock05,hallberg95}
and is evaluated by extending Eq. (\ref{rho3}),
\begin{eqnarray}
|\tilde \psi_r^A\rangle& = &\sum_{i=1}^Me^{-\beta \epsilon^r_i/2}\hat
A|\psi_i^r \rangle \langle \psi_i^r |r\rangle =\hat A|\tilde
\psi_r\rangle,\nonumber \\
\hat \rho_A &\sim &\frac{N_{st}}{R} \sum_{r=1}^R |\tilde \psi_r^A\rangle
\langle \tilde \psi_r^A |.
 \label{eq:psira}
\end{eqnarray}
In the proposed targeting we sum up above contributions with weighting
factors,
\begin{equation}
\hat \rho_{tot} = p_1 \frac{\hat \rho}{\textrm{ Tr} \hat \rho} +
p_2  \frac{\hat \rho_A}{\textrm{ Tr} \hat \rho_A},
\end{equation}
with the restriction $p_1+p_2=1$.  From $\hat \rho_{tot}$ we prepare
the reduced density matrix by integrating out the environment, which
is then used to construct the basis within the infinite and finite
algorithms of the DMRG. \cite{schollwock05} Our way of targeting is in
fact very similar to the one in  Ref. \onlinecite{sota08}, with an additional
random sampling suppressing the non-diagonal terms of $\hat \rho$.  In
such a way we prepare the basis for any $T>0$, whereby limitations are
emerging from the truncation of the basis being more under control for
low $T$. It should also be mentioned that for dynamical response at
particular $\omega$ there is an improvement to target also excited
states corresponding to so called correction
vectors. \cite{kuhner99,sota08} Still, the latter does not affect
quality of the most interesting and challenging regime $\omega \sim 0$
as well as it increases the computational demand, hence we do not
employ it here.

Physical quantities are calculated in the measurement part of the FTD-DMRG
procedure in the same manner as within the FTLM. \cite{jaklic94} A
dynamical autocorrelation function
\begin{equation}
A(\omega)=\frac{1}{Z}\sum_n e^{-\beta E_n} \langle n |\hat A ^\dagger
\frac{1}{\omega -(\hat H -E_n)+i\eta}\hat A|n\rangle, \label{aom}
\end{equation}
is evaluated with the use of two Lanczos series of eigenstates and
eigenenergies,
\begin{eqnarray}
A(\omega)&\approx& \frac{N_{st}}{Z R}\sum_{r=1}^R \sum_{i,j=1}^M
e^{-\beta \epsilon^r_i} \frac{1}{\omega -(\epsilon^{Ar}_j -\epsilon^r_i)+i\eta}
\times \nonumber \\
&&\langle r|\psi_i^r\rangle \langle \psi_i^r |\hat A ^\dagger
|\psi_j^{Ar}\rangle\langle \psi_j^{Ar}|\hat A|r\rangle. \label{aomlm}
\end{eqnarray}
The second Lanczos series of eigenstates $|\psi_j^{Ar}\rangle$ and
eigenenergies $\epsilon_j^{Ar}$ is obtained from second Lanczos
procedure starting from the initial vector $\hat A |r\rangle$.

\section{Model and results}
\label{sec:model}

As a nontrivial test of the method we analyse the dynamics of the 1D
anisotropic Heisenberg model,
\begin{equation}
\hat H=J\sum_{i=1}^L \bigl[\frac{1}{2}( S^+_i S^-_{i+1}+ S^-_i S^+_{i+1} ) +
     \Delta  S^z_i S^z_{i+1} \bigr], \label{ham}
\end{equation}
where $S^\pm_i, S^z_i$ are local spin $S=1/2$ operators, $L$ is the
chain length, $J$ is the exchange coupling (in the following we use
$J=1$) and $\Delta$ the anisotropy parameter. In our calculations we
focus on systems in the absence of the magnetic field, hence on the
subspace $S^z_{tot}= 0$.  As the quantity of interest we choose the
dynamical spin structure factor $S(q,\omega)$ and the corresponding
susceptibility $\chi(q,\omega)$,
\begin{eqnarray}
S(q,\omega)&=&\frac{1}{2\pi}\int_{-\infty}^{+\infty} dt
e^{i\omega t} \langle S^z(q,t) S^z(q,0) \rangle, \nonumber \\
\chi''(q,\omega)&=&\pi(1-e^{-\beta\omega})S(q,\omega). \label{chi}
\end{eqnarray}
As usual within the DMRG technique more accurate results are obtained
with open 
boundary conditions,  \cite{schollwock05} hence one defines
$S^z(q)=\sqrt{2/(L+1)} \sum \sin(qi) S^z_i$ whereby $q=\pi j/(L+1)$
with $j=1, \ldots L$. In our calculations we concentrate on most challenging
$q=Q=\pi$, i.e. $j=L$.

The relaxation function $\Phi(q,\omega)=\chi''(q,\omega)/\omega$ should
be an even function of $\omega$. This represents another nontrivial
test for the FTD-DMRG method.  In addition to considering complete
spectra $\chi''(q,\omega)$ better defined criteria are frequency
moments,
\begin{equation}
 M^{(n)}(q)=\frac{1}{\pi} \int \Phi(q,\omega) \omega^n d\omega.
\end{equation}
Due to symmetry only even $M^n(q)$ are finite while the static
susceptibility corresponds to $\chi^0(q)=M^0(q)$.

In the following we employ the FTD-DMRG method to evaluate
$\Phi(Q,\omega)$ for $\Delta=0,1$ and various $T$.  In the actual
implementation we use the infinite and finite-size DMRG basis
preparation and the calculation of $S(Q,\omega)$ via Eq.(\ref{aomlm})
($A=S_z(Q)$) performed on the system divided into two subblocks of
size $(L-2)/2$ and two coupling sites in between. \cite{schollwock05}
 In the preparatory sweeping typically 1 or 2
sweeps are sufficient for the convergence of the basis.  Important
parameters for the final quality of results are the (subblock) DMRG
truncation number $m$ and the number of Lanczos steps $M$. We are
typically restricted to $m \sim 1000$ and $M\sim 100$. We have two
kinds of sampling. One in the determination of the density matrix
Eq.~(\ref{rho2}), $R=R_1$ for the basis preparation, and the other in
the evaluation of the final Eq.~(\ref{aomlm}), $R=R_2$. While only
modest $R_1 \sim 50$ is adequate, $R_2 \gg 1$ is needed in particular
at low $T$ \cite{jaklic94} to get accurate matrix elements. At higher
$T$ $R_2$ can be reduced effectively to $R_2 \sim 1$. \cite{jaklic94}
 Furtheron we mainly consider $T<0.5$ with $R_2\sim
100$. When evaluating the feasibility of various methods we should
keep in mind that the full exact diagonalization evaluation of
$S(q,\omega)$ at $T>0$ for the model at hand can be performed up to
$L=14-16$, with the FTLM technique up to $L=24$, while in the
following we present the FTD-DMRG results up to $L=40$.

\subsection{XY model}

\begin{figure}[htb] 
\centering 
\epsfig{file=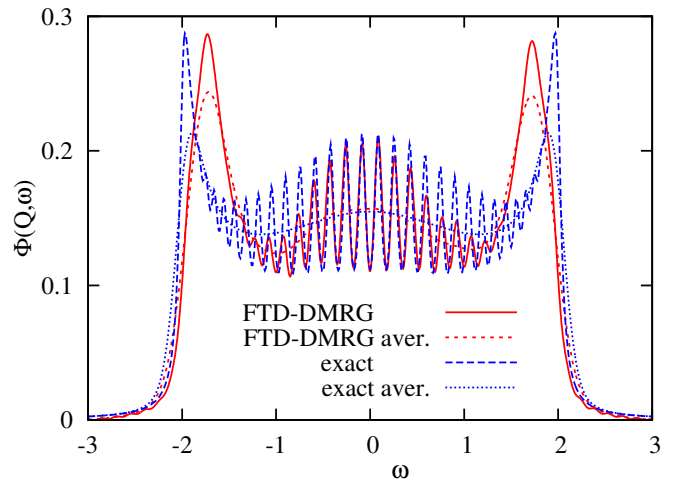,angle=-90,width=86mm}
\caption{(color online) Relaxation function $\Phi(Q,\omega)$ within
  the XY model for $T=0.25$ and a system of $L=36$ sites.  For
  comparison the exact grandcanonical result for spinless fermions is
  shown and the corresponding smoothed curve relevant for $L \to
  \infty$.}
\label{fig1}
\end{figure}

The $\Delta=0$ case maps onto noninteracting spinless fermions via the
Jordan-Wigner transformation and $S(Q,\omega)$ can be expressed for
any $T>0$ in a Lindhard form. For finite $L$ the only caveat is that
the FTD-DMRG is performed within a canonical systems with fixed
$S^z_{tot}=0$, i.e., with fixed number of fermions $N_e=L/2$ while the
usual (easier) evaluation is within the grandcanonical ensemble.  In
Fig.~\ref{fig1} we present the FTD-DMRG result for (unsymmetrized)
relaxation function $\Phi(Q,\omega)$ at low $T=0.25$. Results are for
$L=36$ where the basis is heavily reduced, i.e., only $5\times10^{-4}$
of all states are retained within the final evaluation. For comparison
we show the exact (grandcanonical) fermionic result for the same
system with open boundary condition and for all presented spectra we
use the damping $\eta=0.05$.  Oscillations are a clear sign of
finite-size system and slowly disappear with increasing $T$ and
$L$. The finite-size effect can be avoided by smoothing with a
Gaussian filter with the width adapted to the frequency $\propto
1/L$. From Fig.~\ref{fig1} it is evident that at low $|\omega|<1$ the
agreement between the FTD-DMRG and the exact result is very
satisfactory. At high $\omega \sim 2$ the FTD-DMRG does not fully
reproduce the sharp spectral edge which could be improved by the
introduction of the correction-vector targeting for $\omega \ne 0$
within the method. \cite{kuhner99,schollwock05,sota08}
 
\begin{figure}[htb] 
\centering 
\epsfig{file=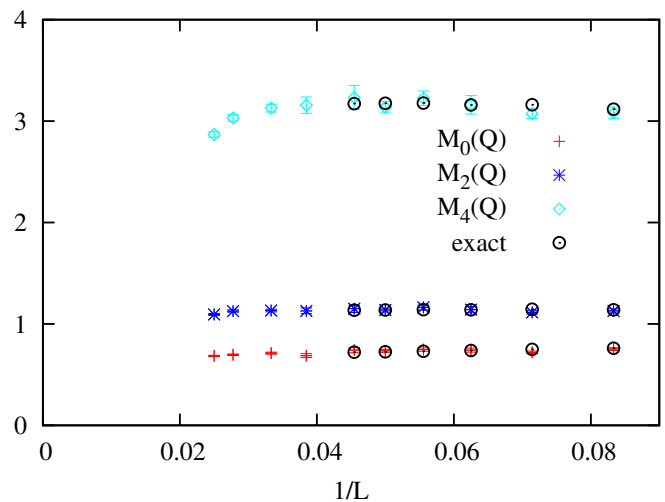,angle=-90,width=86mm}
\caption{(color online) Lowest frequency moments $M_n(Q)$ vs. $1/L$
  for $T=0.25$. For comparison exact moments are shown up to $L=22$.}
\label{fig2}
\end{figure}

Fig.~\ref{fig2} shows the corresponding results for the frequency moments
$M_n(Q)$ displayed vs. $1/L$ obtained with the full basis for $L\le
22$ and with the FTD-DMRG method for $L\le 40$. For comparison also
corresponding exact results are shown within the canonical calculation at
$N_e=L/2$. It is evident that $T=0.25$ is already high enough so that
moments are essentially size independent. Also up to $L=40$ FTD-DMRG
results are well stable, at least for lowest $M_0, M_2$, while for
$M_4$ some deviations originate from high-$\omega$ regime and are also
visible 
in Fig.~\ref{fig1}. At the same time, $M_1,M_3 \approx 0$ is
well reproduced as required by the symmetry of $\Phi(Q,\omega)$.

\subsection{Isotropic Heisenberg model}

The isotropic $\Delta=1$ case (at $S^z_{tot}=0$) representing
marginally gapless system is by far more challenging. For $T>0$ there
are no exact results for dynamical quantities. The bosonization
approach provides a form for $S(q,\omega)$ within the low $\omega - T$
regime. \cite{giamarchi89,bocquet01} Relative to the $\Delta=0$ case
the divergence for $\Delta=1$ is stronger and nontrivial. The
isotropic model has been an obvious target for numerical
methods. Static quantities, as the structure factor $S(q)$ and
$\chi^0(q)$ have been evaluated using the quantum Monte Carlo (QMC)
method and the high-$T$ expansion, \cite{starykh97,grossjohann09}
recently also with the time-dependent DMRG, \cite{barthel09} but only
for $q \ne Q$ so far. An obvious deficiency is in results for dynamic
quantities at $\omega \sim 0$ since the QMC approach (due to the
Maximum Entropy procedure) seems to have considerable uncertainty in this
regime. \cite{grossjohann09} On the other hand, the latter regime is
frequently just the most interesting, e.g., in connection with the NMR
relaxation rate $1/T_1 \propto \sum_q A_q S(q,\omega \to 0)$, with
transport quantities etc.

\begin{figure}[htb] 
\centering 
\epsfig{file=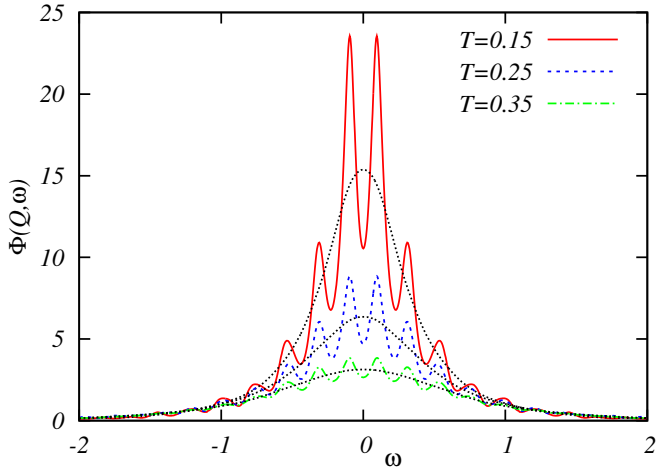,angle=-90,width=86mm}
\caption{(color online) Symmetrized $\Phi(Q,\omega)$ for the isotropic Heisenberg
  model shown for $L=40$ and $T=0.15, 0.25, 0.35$. Presented
  are also finite-size smoothed spectra (dotted line).}
\label{fig3}
\end{figure}

In Fig. \ref{fig3} we present results for $\Phi(Q,\omega)$ obtained
for $L=40$ sites and different $T$. Since spectra are peaked at
$\omega =0$ (in contrast to Fig. \ref{fig1}) finite-size oscillations
are more pronounced. Hence, also smoothed curves (Gaussian width
$\sigma = 4\cos (\pi L/2(L+1))/\sqrt{2}$) are presented as relevant for
$L \to \infty$. We note that such spectra are nearly $L$-independent
($L=16-40$) for $\omega>0.5$ whereas for $\omega \sim 0$ still scale
as $a+b/L$.

\begin{figure}[htb] 
\centering 
\epsfig{file=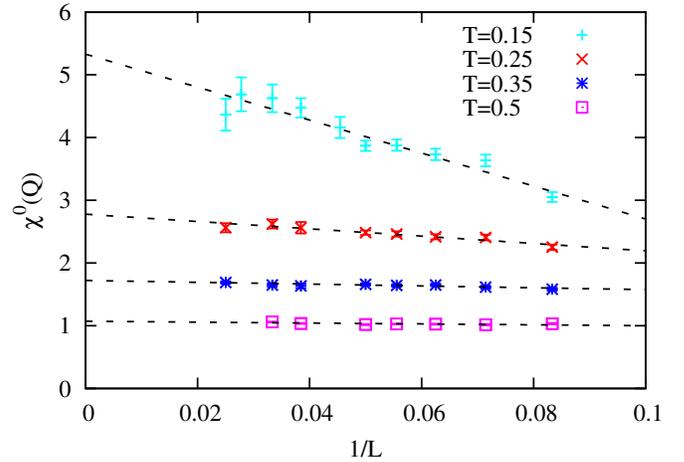,angle=-90,width=86mm}
\caption{(color online) $\chi^0(Q)$ for the isotropic model vs.  $1/L$
for different $T$  as calculated via the FTD-DMRG method for $L=22 - 40$ and via
FTLM for $L=12 - 20$.}
\label{fig4}
\end{figure}

\begin{figure}[htb] 
\centering
\epsfig{file=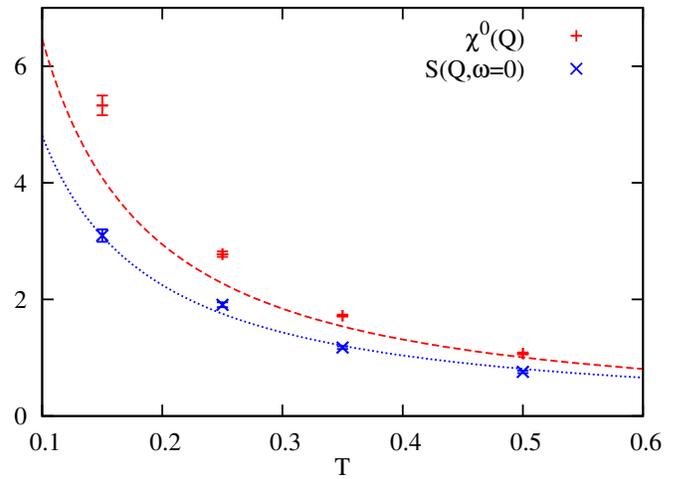,angle=-90,width=86mm}
\caption{(color online) Scaled values for $\chi^0(Q)$ and
  $S(Q,\omega=0)$ vs. $T$ for the isotropic model. The dashed curve
  represents $\chi^0(Q)$ using the analytical form as extracted from
  the QMC results Ref. \onlinecite{starykh97}. Dotted $S(Q, \omega=0)$ curve
  is the fit as deduced from the analytical
  approximation. \cite{bocquet01}}
\label{fig5}
\end{figure}

On the other hand, static $\chi^0(Q)$ can be extracted directly
without invoking any smoothing and FTD-DMRG results combined with the
FTLM results for $L=12-20$ are shown in Fig.~\ref{fig4} scaled
vs. $1/L$. Deviations from the linear scaling mostly emerge from the
random sampling in the basis preparation and the dynamical quantity
evaluation, and for the latter are indicated with error bars.  Final
scaled FTD-DMRG results for $\chi^0(Q)$ vs. $T$ are shown in
Fig.~\ref{fig5}, together with the result of the QMC analysis
\cite{starykh97} of the analytical expression
\begin{equation}
\chi^0(Q)=\frac{a}{T}[\ln(b/T)]^{1/2}.
\end{equation}
Our FTD-DMRG result is quite
consistent with QMC results at higher $T>0.3$. Still it is indicative
that we get higher values (beyond error bars) for $T<0.3$.

Finally, we present in the same Fig.~\ref{fig5} also scaled values of
$S(Q,\omega=0)$ vs. $T$. Bosonization theory gives \cite{bocquet01}
\begin{equation}
S(Q,0)= \frac{A}{T}[\ln(\Lambda/T)]^{1/2}
\end{equation}
 also fitted to our results with
$\Lambda=24.27$ taken from Ref. \onlinecite{barzykin01} and adjusted
$A\sim0.205$.  The agreement with the analytical fit is very good
although there seems to be substantial difference in the prefactor
$A$. \cite{bocquet01} On the other hand, it should be reminded that
for this quantity there are no reliable larger-system alternative
results since the QMC analysis \cite{starykh97,grossjohann09} appears
to have some difficulties in the regime $\omega \sim 0$.

\section{Conclusions}
\label{sec:conclusions}

In conclusion, we have introduced the FTD-DMRG method, which is the
extension of the density matrix-based optimization of target states
and the FTLM method for the evaluation of dynamical quantities at
$T>0$.  It is so far well founded and tested for relatively low T and
not too large systems, e.g., $L < 40$, while the feasibility or
possible breakdown at larger $T$ should still be understood.
Presented results are obtained for systems with $\tilde Z <200$
(normalized so that $\tilde Z(T=0)=1$) although the method is not in
principle limited to low $T$ since it is not essential that all
relevant many-body states are well represented, in analogy to the
FTLM. \cite{jaklic94} The emphasis so far is on the most challenging
$\omega \sim 0$ dynamical response while higher $\omega$ could be
improved by extending the density matrix by optimizing the correction
vector at particular $\omega$. \cite{kuhner99,sota08} As the test we
use the $\Delta=0$ case which is nontrivial for the FTD-DMRG method
while exact results are available via the spinless-fermion
representation. On the other hand, results for the isotropic
$\Delta=1$ case where we concentrate on the low $\omega - T$ regime of
dynamical spin correlations $S(Q,\omega)$ show that the presented
method goes beyond the capabilities of up-to-date numerical methods,
e.g., in the case of $S(Q,\omega=0)$. Clearly, more effort is needed
to examine in more detail the feasibility of the new method.

\begin{acknowledgments}
We authors acknowledge helpful discussions with T. Tohyama and S. Sota
as well as the support of the Slovenia-Japan Research Cooperative
grant and the Slovenian Agency grant No. P1-0044. 
\end{acknowledgments}


\end{document}